\documentclass[aps,preprintnumbers,amsmath,amssymb,eqsecnum,twocolumn]{revtex4}

\usepackage{graphicx}

\begin{document}

\title{Theoretical model for the structural phase transition at the metal-insulator transition in polymerized KC$_{60}$}
\author{B.\ Verberck$^{1}$, A.V. Nikolaev$^{1,2}$, and K.H. Michel$^{1}$}
\affiliation{$^1$Department of Physics, University of Antwerp, Universitaire Instelling Antwerpen, Universiteitsplein 1, 2610 Antwerp, Belgium\\
$^2$Institute of Physical Chemistry of the Russian Academy of Sciences, Leninskii prospect 31, 117915, Moscow, Russia
}
\date{\today}

\begin{abstract}
The recently discovered structural transition in polymerized KC$_{60}$ at about 50 K results in a doubling of the
unit cell volume and accompanies the metal-insulator transition.  Here we show that the
$(\vec{a}+\vec{c},\vec{b},\vec{a}-\vec{c})$ superstructure results from small orientational charge density waves
along the polymer chains and concomitant displacements of the surrounding K$^+$ ions.  The effect is
specific for the space group $Pmnn$ of KC$_{60}$ and is absent in Rb- and CsC$_{60}$ (space group $I2/m$).  The
mechanism is relevant for the metal-insulator transition.
\end{abstract}

\maketitle

\section{Introduction}
\begin{figure} 
\includegraphics{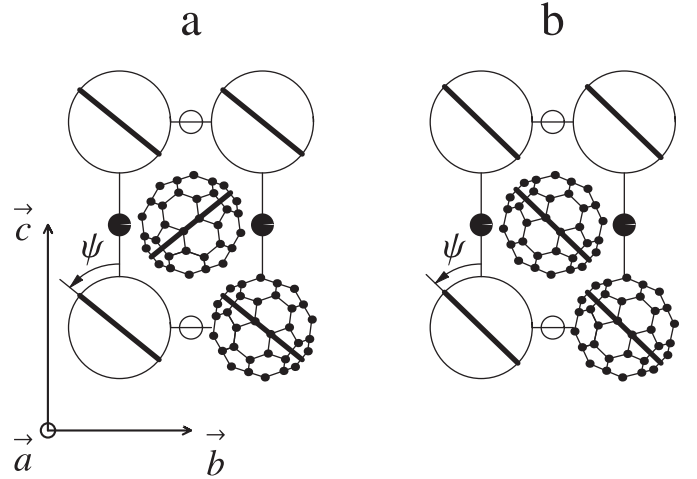} 
\caption{
Crystal structures projected onto the crystallographic $(\vec{b},\vec{c})$ plane: (a) $Pmnn$, (b) $I2/m$. The thick bars represent the
projection of the cycloaddition planes.  Polymerization occurs along $\vec{a}$.  The alkali-metals located in $(\vec{b},\vec{c})$ planes and at
$\pm \vec{a}/2$ are denoted by full and empty circles.
} 
\label{figurestructures} 
\end{figure}

Among the alkali-metal-doped fullerides A$_x$C$_{60}$, where A=K, Rb, Cs, the $x=1$ compounds \cite{Win} have
attracted attention, because they form plastic cubic crystalline phases with rock salt structure at high
temperature ($T>350$ K) and stable polymeric phases \cite{Pek} of reduced symmetry at lower $T$.  In addition, via
rapid quenching, a metastable dimer phase was obtained \cite{Zhu,Osz}.  It was suggested that the dimer structure
should be a Peierls insulator \cite{Zhu}.  In the following we will restrict
ourselves to the polymer phases.  There, the C$_{60}$ molecules are linked through a [2+2]
cycloaddition and form chains along the
former cubic $[110]$ direction.  The orientation of the polymer chains is characterized by the angle $\psi$ of
the planes of cycloaddition with the former cubic $[001]$ direction.  The structure of KC$_{60}$ is
orthorhombic \cite{Ste,Lau1} [space group $Pmnn$, Fig.\ \ref{figurestructures}(a)]; the orthorhombic $\vec{a}$
axis is parallel to the axis of polymerization and the $\vec{c}$ axis parallel to the former cubic $[001]$
direction.  The $Pmnn$ structure is characterized by alternating orientations $\pm\psi$ of the polymer chains
in successive $(\vec{a},\vec{b})$ planes and the same orientation of the chains within one plane,
$|\psi|\approx 45^\circ$.  On the other hand, the structure of both polymerized RbC$_{60}$ \cite{Lau1} and
CsC$_{60}$ \cite{Rou} is monoclinic, space group $I2/m$ [Fig.\ \ref{figurestructures}(b)].  Here the polymer
chains have the same orientation $\psi$ not only within the same $(\vec{a},\vec{b})$ plane but also in successive
(001) planes.

In addition to the structural differences, the electronic and magnetic properties of KC$_{60}$ on one hand and
Rb-, CsC$_{60}$, on the other hand, are different (for a review, see Ref.\ \onlinecite{For}).  Neither the
differences in structure nor those in electromagnetic properties can be simply related to the relatively small
differences in the lattice constants among the AC$_{60}$ compounds.  We then conclude that more
alkali-metal-specific effects are relevant.  To explain the structural differences of the polymer phases of KC$_{60}$ (space group
$Pmnn$) and Rb- and CsC$_{60}$ (space group $I2/m$), it has turned out that the quadrupolar polarizability of the
alkali-metal ions is the decisive alkali-specific-characteristic \cite{Mic,Ver}.  The quadrupolar polarizability of the
alkali-metal ion A$^+$ is related to the average radius $d_{\rm A}$ of its first valence electron $d$ shell.  Indeed,
$d_{\rm Rb}=1.82\mbox{ \AA}$ and $d_{\rm Cs}=1.87\mbox{ \AA}$ are close to each other but quite different from
$d_{\rm K}=1.47\mbox{ \AA}$.  The partial occupancy of the first excited $d$ state in the solids is possible
because of the large interstitial space available for the alkali-metals, this relatively large space being a unique
property of the AC$_{60}$ compounds.

Recently, a combined study of the electronic and structural properties of KC$_{60}$ has been carried out by
Coulon {\em et al.\ }\cite{Cou}: x-ray
diffraction studies have revealed a structural phase transition in polymerized KC$_{60}$ at 60 K $\leq T_c \leq$ 65 K, while ESR measurements
have shown the existence of a metal-insulator transition at $T\approx$ 50 K, which stresses once again the close connection between
structural and electronic properties in the AC$_{60}$ alkali-metal fullerides.  Above the structural critical temperature $T_c$, the Bravais lattice
of polymerized KC$_{60}$ is orthorhombic; at $T_c$, the crystal changes to a $(\vec{a}+\vec{c},\vec{b},\vec{a}-\vec{c})$
superstructure, which can be viewed as a doubling of the basis vectors $\vec{a}$ and $\vec{c}$.  Concerning the possible nature of this
structural phase transition (i.e., the doubling mechanism), it was pointed out by Coulon {\em et al.\ }\cite{Cou} (i) that displacements of the
C$_{60}$ centers of mass can be ruled out and (ii) that a combination of a charge modulation on the C$_{60}$ monomers and large correlated K
displacements is an appealing hypothesis.

We note that such a combined structural and metal-insulator phase transition has not yet been observed in
RbC$_{60}$ \cite{Lau}.  Bearing in mind the structural differences of the polymer phases of KC$_{60}$ and
RbC$_{60}$, the presence/absence of a second structural phase transition in AC$_{60}$ seems to be related to the
precise orientations of the polymer chains in $Pmnn$ and $I2/m$, respectively.

Here we propose a doubling mechanism that accounts for the observed structural phase transition in KC$_{60}$ on
one hand, and for the absence of such a transition in RbC$_{60}$ on the other hand, both in agreement with the
present experimental knowledge of these compounds.

\section{Doubling Mechanism}\label{doubling}
We consider rigid KC$_{60}$ and RbC$_{60}$ crystals in the polymerized phase.  For KC$_{60}$, this phase has space group $Pmnn$, while for
RbC$_{60}$ the space group is $I2/m$.  The $I2/m$ space group has a monoclinic Bravais lattice; we will treat it
as an orthorhombic lattice,
however, since the ``monoclinic" angle $\alpha=90.316^\circ$ of RbC$_{60}$ obtained by neutron scattering measurements \cite{Huq} is extremely
close to $90^\circ$.  The centers of mass of the $N$ C$_{60}$ monomers are then located on the lattice points of a body-centered orthorhombic
lattice with basis vectors $\vec{a}=a\vec{e}_X$, $\vec{b}=b\vec{e}_Y$, and $\vec{c}=c\vec{e}_Z$, where $(\vec{e}_X,\vec{e}_Y,\vec{e}_Z)$ are
the basis vectors of the underlying Cartesian coordinate system.  The axes of polymerization are parallel to $\vec{a}$.  A lattice point is
labeled by indices $\vec{n}=(n_1,n_2,n_3)$, which are either all integer numbers or all integer numbers
$+\frac{1}{2}$, corresponding, respectively, to the corner and the center points of the unit cells.  The position vector of lattice point $\vec{n}$ reads
\begin{equation}
\vec{X}({\vec{n}})=n_1\vec{a}+n_2\vec{b}+n_3\vec{c}.
\end{equation}
To each polymer chain, a rotation angle $\psi$ can be assigned.  We let $\psi=0$ correspond to the situation where the polymer chain is in the
standard orientation, which is defined as the orientation where the plane of cycloaddition is parallel to the $(\vec{a},\vec{c})$ plane.  The
angle $\psi$ then measures a counterclockwise rotation of the polymer chain about $\vec{a}$.  More generally, the orientation angle can be
seen as a property of a single C$_{60}$ monomer; therefore we write $\psi\equiv\psi(\vec{n})$.  However, since
all the C$_{60}$ monomers in the
same polymer chain have the same orientation angle and since a polymer chain can be addressed by the indices $(n_2,n_3)$, $\psi$ is
independent of the index $n_1$.  Furthermore, the $Pmnn$ structure is characterized by an alternation of the orientations of the polymer
chains along the $\vec{c}$ axis only; hence for KC$_{60}$ $\psi$ is also independent of the index $n_2$:
\begin{equation}
\psi(\vec{n})\equiv\psi(n_3)=(-1)^{2n_3}\psi_{\text{KC}_{60}}. \label{psiKC60}
\end{equation}
The structure $I2/m$ is characterized by an equal orientation of all the polymer chains in the
crystal.  Therefore, one has for RbC$_{60}$:
\begin{equation}
\psi(\vec{n})\equiv\psi_{\text{RbC}_{60}}.
\end{equation}
The chain orientation angles have been redetermined recently by neutron scattering experiments \cite{Huq}:
$\psi_{\text{KC}_{60}}=50^\circ$, $\psi_{\text{RbC}_{60}}=46^\circ$.

Due to the charge transfer of one electron from an alkali-metal atom to a C$_{60}$ molecule, a charge distribution $\rho$ with total charge $-e$
exists on every C$_{60}$ monomer.  For a given C$_{60}$ monomer $\vec{n}$, we introduce a Cartesian coordinate system with basis vectors
$(\vec{e}_x,\vec{e}_y,\vec{e}_z)$ and the center of mass of the monomer as origin.  The charge distribution can then be written as
\begin{equation}
\rho\equiv\rho(\vec{r};\psi(\vec{n}))=\rho(x,y,z;\psi(\vec{n})),
\end{equation}
where $\vec{r}=x\vec{e}_x+y\vec{e}_y+z\vec{e}_z$.

\begin{figure*} 
\includegraphics{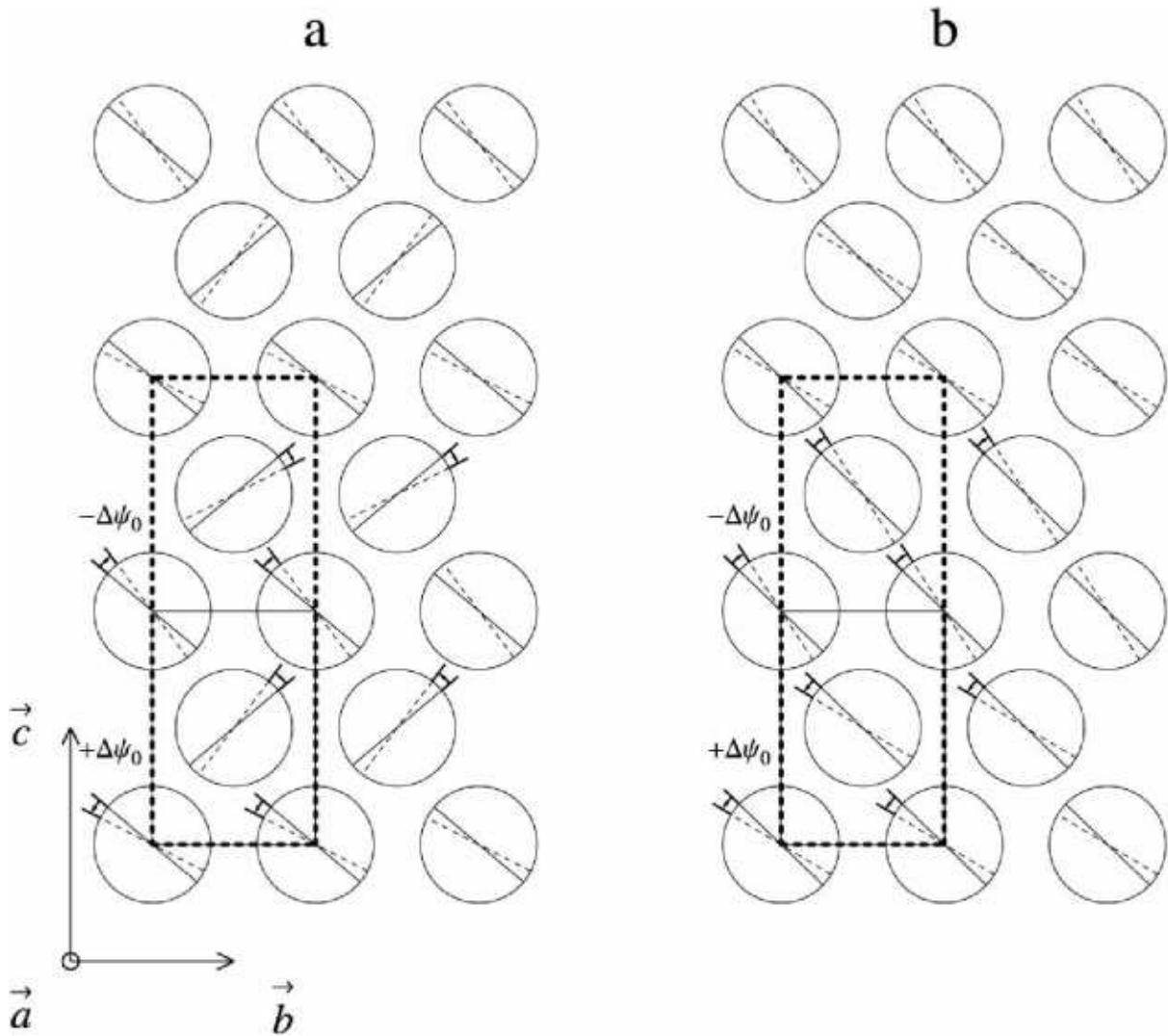} 
\caption{
Illustration of the proposed doubling mechanism.  Shown is a projection of the C$_{60}$ molecules, symbolized
by circles, onto the $(\vec{b},\vec{c})$ crystallographic plane (compare with Fig.\ \ref{figurestructures}).
The alkali-metal atoms have been omitted for clarity.  The solid and dashed lines in the circles represent the
orientations of the charge distributions: (a) KC$_{60}$, (b) RbC$_{60}$.  The solid lines correspond to the
original $Pmnn$ (a) and $I2/m$ (b) structures, the suggested doubling mechanism consists of angular deviations
of the electronic density distributions as indicated by the dashed lines.  The proposed scheme results in a
doubling of the lattice vector $\vec{c}$, while the lattice vector $\vec{b}$ remains the same.  The original
(projection of the) unit cell is shown as a solid frame, the (projection of the) unit cell taking into account
the electronic deviations is shown as a dashed frame.  Along the $\vec{a}$ axis, a doubling pattern (angular
deviations of the charge distributions) similar to the doubling along the $\vec{c}$ axis occurs.
} 
\label{fig2} 
\end{figure*}

As a mechanism for the doubling of the lattice basis vectors $\vec{a}$ and $\vec{c}$, we suggest the following: while retaining the rigid
structure for the {\em C nuclei and the closed $\pi$- and $\sigma$-shell electrons}, small orientational
deviations of the {\em valence
electron density $\rho$} on every C$_{60}$ monomer from this structure are allowed to occur, in such a way that these deviations (and
therefore equivalent lattice points) have periodicities $2\vec{a}$, $\vec{b}$, and $2\vec{c}$.  A rotation of the electron density can occur
because there are orbital degrees of freedom for one valence electron on the threefold degenerate $t_{1u}$ lowest unoccupied molecular
orbital (LUMO) level \cite{Had}.  The valence electron density depends on the coefficients of expansion in terms of these three orbitals and
this leads to its effective rotation.  We will limit ourselves to rotations of the charge distributions about the polymer chain axes.  To
denote the angular deviation of the charge distribution from $\psi(\vec{n})$, we use the notation $\Delta\psi(\vec{n})$:
\begin{equation}
\rho\equiv\rho(\vec{r};\psi(\vec{n})+\Delta\psi(\vec{n}))=\rho(x,y,z;\psi(\vec{n})+\Delta\psi(\vec{n})).
\end{equation}
In order to have the desired periodicity changes, we take
\begin{eqnarray}
\Delta\psi(n_1,n_2,n_3) & = & \Delta\psi(n_1+\frac{1}{2},n_2+\frac{1}{2},n_3+\frac{1}{2}) \nonumber \\
 & = & \left\{
	\begin{array}{l}
	+\Delta\psi_0\mbox{ if $n_1+n_3$ even} \\
	-\Delta\psi_0\mbox{ if $n_1+n_3$ odd},
	\end{array}
	\right. \label{Deltapsi0ref}
\end{eqnarray}
with $n_1,n_2,n_3\in\mathbb{Z}$.  In Eq.\ (\ref{Deltapsi0ref}), $\Delta\psi_0$ represents the angle measuring the deviation from $\psi$.  The
mechanism is illustrated schematically in Fig.\ \ref{fig2}.  In Sec.\ \ref{charge} we will comment on the charge distribution
model to be used.

It is convenient to introduce spherical coordinates $(r,\theta,\phi)$, related to the Cartesian coordinates $(x,y,z)$ via
\begin{equation}
\begin{array}{c}
y=r\sin\theta\cos\phi, \\
z=r\sin\theta\sin\phi, \\
x=r\cos\theta, \\
\end{array}
\label{cotrafo}
\end{equation}
since a counterclockwise rotation of the charge distribution $\rho$ over an angle $\psi+\Delta\psi$ about the $\vec{a}$ axis is then simply
achieved by replacing $\phi$ by $\phi-\psi-\Delta\psi$:
\begin{equation}
\rho(r,\theta,\phi;\psi(\vec{n})+\Delta\psi(\vec{n}))=\rho(r,\theta,\phi-\psi(\vec{n})-\Delta\psi(\vec{n});0).
\end{equation}
We simplify the notation and write from now on:
\begin{equation}
\rho(r,\theta,\phi)\equiv\rho(r,\theta,\phi;0).
\end{equation}

\section{Potential Energy}\label{potentialenergy}
In order to determine whether the above-described electronic density distortions can occur or not, it is necessary to investigate the
potential energy of the crystal.  We consider the contribution ${\cal U}$ to the potential energy arising from C$_{60}$--C$_{60}$
interactions, which can be written as a sum of pair potentials:
\begin{equation}
{\cal U}=\frac{1}{2}\sum_{\vec{n},\vec{n}'}U(\vec{n},\vec{n}'). \label{U}
\end{equation}
Since we are interested in how the potential energy of the crystal is influenced by the above-described orientational deviations
$\Delta\psi(\vec{n})$ of the charge distributions on all C$_{60}^-$ monomers, only electronic interactions have to be taken into account (van
der Waals-type interactions refer to the neutral C cores, which in our model are assumed to constitute a rigid structure).  The electrostatic
energy of two C$_{60}^-$ monomers at lattice points $\vec{n}$ and $\vec{n}'$ is given by
\begin{subequations}
\begin{equation}
U(\vec{n},\vec{n}')=\frac{1}{4\pi\epsilon_0}\int d\vec{r}\int d\vec{r}'\frac{\rho(\vec{r};\alpha(\vec{n}))\rho(\vec{r}';\alpha(\vec{n}'))}
{|\vec{r}-\vec{r}'-\vec{X}(\vec{n}'-\vec{n})|},
\label{Unnprime}
\end{equation}
where we have introduced the notation
\begin{equation}
\alpha(\vec{n})\equiv\psi(\vec{n})+\Delta\psi(\vec{n}).
\end{equation}
\end{subequations}
The integration variables $\vec{r}$ and $\vec{r}'$ in Eq.\ (\ref{Unnprime}) refer to the {\em local} coordinate systems associated with the
respective lattice sites $\vec{n}$ and $\vec{n}'$; hence the appearance of the relative position vector
$\vec{X}(\vec{n}'-\vec{n})=\vec{X}(\vec{n}')-\vec{X}(\vec{n})$.  Using the previously introduced spherical coordinates, $U(\vec{n},\vec{n}')$
can be rewritten as
\begin{widetext}
\begin{subequations}
\begin{eqnarray}
U(\vec{n},\vec{n}') & = & \frac{1}{4\pi\epsilon_0}\int_0^{\infty} r^2dr\int_0^{\pi}\sin\theta d\theta\int_0^{2\pi} d\phi
\int_0^{\infty} r'^2dr'\int_0^{\pi}\sin\theta 'd\theta'\int_0^{2\pi} d\phi' \nonumber \\
 &  & \times\frac{\rho(r,\theta,\phi-\psi(\vec{n})-\Delta\psi(\vec{n}))\rho(r',\theta',\phi'-\psi(\vec{n}')-\Delta\psi(\vec{n}'))}
 {|\vec{r}-\vec{r}'-\vec{X}(\vec{n}'-\vec{n})|}, \label{Unnprime2a} \\
|\vec{r}-\vec{r}'-\vec{X}(\vec{n}'-\vec{n})| & = & \left\{[r\cos\theta-r'\cos\theta'-(n'_1-n_1)a]^2
+[r\sin\theta\cos\phi-r'\sin\theta'\cos\phi'-(n'_2-n_2)b]^2 \right. \nonumber \\
 &  & \left.+[r\sin\theta\sin\phi-r'\sin\theta'\sin\phi'-(n'_3-n_3)c]^2\right\}^{1/2}. \label{Unnprime2b}
\end{eqnarray}
\end{subequations}
\end{widetext}

As will be discussed in more detail in Sec.\ \ref{charge}, the charge distribution $\rho$ can be expanded in even
multipoles $l=0,2,4,\hdots$\ .
The lowest-order term containing angular dependence will therefore be a monopole-quadrupole interaction.  As a consequence, the angular
dependent part of the potential energy $U$, being of the Coulomb-type, decreases fast enough with the distance, and it is justified to
consider a limited number of nearest-neighbor interactions.  We write Eq.\ (\ref{U}) as
\begin{equation}
{\cal U}=\frac{1}{2}\sum_{\vec{n}}V(\vec{n}), \label{U2}
\end{equation}
where
\begin{equation}
V(\vec{n})=\sum_{\vec{\mu}}U(\vec{n},\vec{n}+\vec{\mu}).
\end{equation}
The summation over $\vec{\mu}$ runs over the fourteen nearest-neighbor sites with relative indices $\vec{\mu}=(\pm 1,0,0)$, $(0,\pm 1,0)$,
$(0,0,\pm 1)$, $(\pm\frac{1}{2},\pm\frac{1}{2},\pm\frac{1}{2})$.  An analysis of the occuring orientations of the charge distributions on the
various C$_{60}^-$ monomers reveals that only two different types of lattice sites exist: if we let $n_1$, $n_2$, and $n_3$ be integers, then
all sites with indices $(n_1,n_2,n_3)$ or $(n_1+\frac{1}{2},n_2+\frac{1}{2},n_3+\frac{1}{2})$, satisfying $n_1+n_3$ even, have an equivalent
neighborhood.  We call these sites type I sites.  The other sites ($n_1+n_3$ odd), which we label as type II sites, have also an equivalent
environment, but one that is different from the type I neighborhood.  We summarize this observation by writing
\begin{eqnarray}
V(n_1,n_2,n_3) & = & V(n_1+\frac{1}{2},n_2+\frac{1}{2},n_3+\frac{1}{2}) \nonumber \\
 & = & \left\{
	\begin{array}{l}
	V^{\text{I}}\mbox{ if $n_1+n_3$ even} \\
	V^{\text{II}}\mbox{ if $n_1+n_3$ odd},
	\end{array}
	\right. \label{VIandVII}
\end{eqnarray}
with $n_1,n_2,n_3\in\mathbb{Z}$.  Note that Eq.\ (\ref{VIandVII}) is consistent with the imposed periodicity conditions (\ref{Deltapsi0ref}).
The potential energy of the entire crystal due to all electrostatic C$_{60}^-$--C$_{60}^-$ interactions is obtained by carrying out the
summation of Eq.\ (\ref{U2}), which runs over all lattice points:
\begin{equation}
{\cal U}=\frac{N}{4}(V^{\text{I}}+V^{\text{II}}).
\end{equation}
The functions $V^{\text{I}}$ and $V^{\text{II}}$ depend only on the angle $\Delta\psi_0$, introduced in Eq.\ (\ref{Deltapsi0ref}), which is a
measure for the deviation from the undistorted structure.  To emphasize this dependence, we write
\begin{equation}
{\cal U}\equiv{\cal U}(\Delta\psi_0).
\end{equation}

\section{Charge Distribution}\label{charge}
The key point in the proposed model is the expression for the electronic density $\rho(\vec{r})$ of a C$_{60}^-$ unit in a polymer chain.
Polymerization leads to a reduction of the symmetry (in comparison with the icosahedral symmetry of C$_{60}$) and the charge distribution can
be expanded in multipoles with $l=0,2,4,\hdots$\ .  In earlier theoretical work \cite{Mic,Ver}, a quadrupolar
charge distribution model was used in explaining the structural phase transition from the cubic (unpolymerized) to the orthorhombic (polymerized) phase of the AC$_{60}$
alkali-metal fullerides.  By using a simplified Slater-Koster tight-binding approach to determine $\rho(\vec{r})$, it has been shown \cite{Nik} that
such a quadrupolar model is a reasonable first approximation.  (Due to the $D_{2h}$ symmetry of a polymer chain, only even multipoles occur.)
However, by experimenting with various quadrupolarlike electronic densities, we find that the energy ${\cal U}$ depends very sensitively on
the precise location of the point charges used in constructing the charge distribution and that here no conclusion can be drawn based upon
quadrupolar models.  In our view, it is necessary to go beyond the quadrupolar contribution and to take into account higher multipoles.
Indeed, if one examines the expansion of the angular part $\rho_a(\theta,\phi)$ of the charge distribution in terms of spherical harmonics,
calculated in Ref.\ \onlinecite{Nik}, one sees the relevance of the higher-order terms.  In particular, the $(l=8,m=8)$ term seems to play an
important role.  The importance of relatively-high-order multipoles is not surprising: for example, orientational ordering in solid C$_{60}$
(fullerite) is described using molecular and site-symmetry-adapted functions (SAFs) belonging to the manifolds $l=6,10,12$ \cite{Mic2}.
Furthermore, in expanding the van der Waals interactions between polymers in AC$_{60}$ in terms of SAFs, we have recently discovered a
similar pattern: the term with $(l=8,m=8)$ is remarkably dominant.

\begin{figure*}
\includegraphics{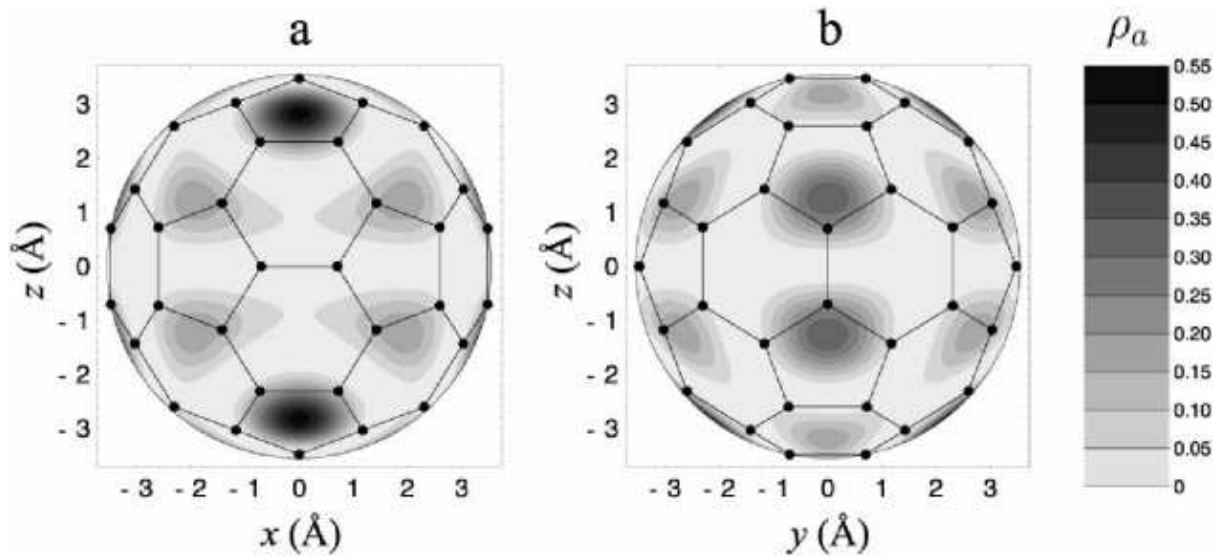} 
\caption{
Angular part $\rho_a(\theta,\phi)$ of the charge distribution of a C$_{60}^-$ monomer in the standard orientation,
plotted as a contourplot on a sphere with radius $R=\mbox{3.55 {\AA}}$, together with the C$_{60}$ cage.  (a)
Projection of $\rho_a$ onto the $(\vec{a},\vec{c})$ plane.  (b) Projection of $\rho_a$ onto the
$(\vec{b},\vec{c})$ plane.  The coordinates $(x,y,z)$ and $(\theta,\phi)$ are related to each other via
Eq.\ (\ref{cotrafo}) and the constraint $\sqrt{x^2+y^2+z^2}=R$.  On both projections, one can see clearly that the
charge is mainly concentrated in the equatorial region ($x=0$).  Note the local maxima in the charge density near
the C atoms participating in the cycloaddition bonds, and the four absolute maxima located at the centers of
pentagons (electron-poor regions of the neutral C$_{60}$ molecule).
} 
\label{figurechargedistribution} 
\end{figure*}

These observations lead us to use the electronic density of Ref.\ \onlinecite{Nik}, which can be written as 
\begin{subequations}
\begin{equation}
\rho(r,\theta,\phi)=-e\rho_r(r)\rho_a(\theta,\phi), \label{rho}
\end{equation}
where the charge of the electron $-e$ has been factored out in order to ensure that the angular part $\rho_a$ is dimensionless.  The radial
part $\rho_r$ is not relevant for our purposes; we locate the charge on a sphere with radius $R=3.55$ {\AA}, which is the radius of a C$_{60}$
molecule \cite{Dav}:
\begin{equation}
\rho_r(r)=\frac{\delta(r-R)}{R^2}. \label{rhor}
\end{equation}
If a more refined model for $\rho_r$ is used, the main effect is a renormalization of the intersite interactions.  On the other hand, the
angular part $\rho_a$ is anisotropic and results in multipole interactions of different sign and magnitude.  It is given by
\begin{eqnarray}
\rho_a(\theta,\phi) & = & [-0.70699Y_5^{1,s}(\theta,\phi)+0.30659Y_5^{3,s}(\theta,\phi) \nonumber \\
 & & +0.63731Y_5^{5,s}(\theta,\phi)]^2. \label{rhoa}
\end{eqnarray}
\end{subequations}
In Eq.\ (\ref{rhoa}), $\rho_a$ is written as the modulus squared of the wave function $\psi_3(\theta,\phi)$, which
is one of the three degenerate $t_{1u}$ functions of a C$_{60}^-$ ion \cite{Nik2}.  The real functions
$Y_l^{m,s}(\theta,\phi)$ are the sine spherical harmonics defined in Ref.\ \onlinecite{Bra}.  Note that
Eq.\ (\ref{rhoa}) is exact within the framework of the tight-binding approach.  In
Fig.\ \ref{figurechargedistribution}, the angular electron density $\rho_a$ for a C$_{60}^-$ ion in the standard
orientation is shown, projected onto the $(\vec{b},\vec{c})$ and $(\vec{a},\vec{c})$ planes.  The charge is mainly
concentrated in the equatorial region ($x=0$ or, equivalently, $\theta=\pi/2$), in agreement with recent NMR
results \cite{Swi}.  As can be seen clearly in Fig.\ \ref{figurechargedistribution}(a), the four absolute maxima of
the charge distribution $\rho_a$ coincide with the centers of pentagons \cite{Lau}, which is very reasonable.
Indeed, it is known that in the neutral molecule, the centers of pentagons are the electron-poor regions and any
additional negative charge experiences minimal repulsion at these centers.  A direct consequence of this fact is
the $Pa\overline{3}$ phase of solid C$_{60}$: here, an electron-rich region of one molecule (double C--C bond)
faces a pentagon of its neighbor \cite{Dav2}.

\section{Potential Energy Calculations}
\begin{figure} 
\includegraphics{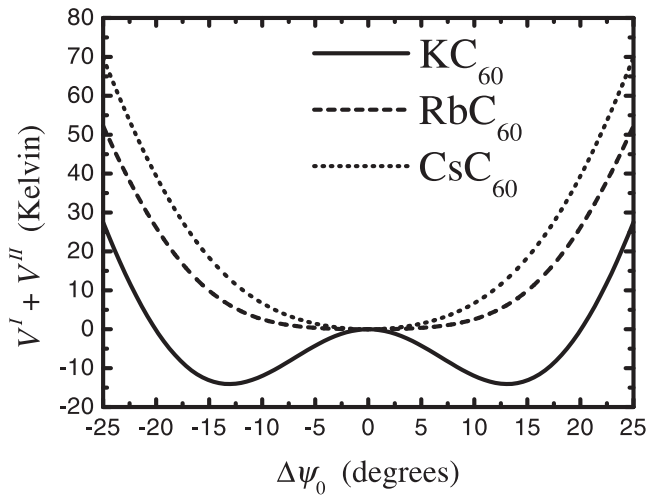} 
\caption{
Dependence of the potential energy $V^I+V^{II}$ (units of kelvin), to which the total potential energy of the crystal due to
C$_{60}^-$--C$_{60}^-$ interactions is proportional, on the deviation angle $\Delta\psi_0$ (units of degrees).  The energy scale has been shifted
for all curves so that the undistorted structure ($\Delta\psi_0=0^\circ$) corresponds to zero potential energy.  The double minimum
for KC$_{60}$ (solid line) shows that an energetically more favorable configuration than the $Pmnn$ structure exists, featuring rotated
electronic densities, while for RbC$_{60}$ (dashed line) and CsC$_{60}$ (dotted line) deviations of the electronic distributions on the
C$_{60}^-$ monomers from the original $I2/m$ structure do not lower the potential energy.
} 
\label{figureenergy} 
\end{figure}

The potential energy $U$ of two interacting C$_{60}^-$ monomers can now be written down explicitly.  Combining Eqs.\
(\ref{Unnprime2a}), (\ref{Unnprime2b}), (\ref{rho}), and (\ref{rhor}), one gets
\begin{widetext}
\begin{subequations}
\begin{eqnarray}
U(\vec{n},\vec{n}') & = & F\int_0^{\pi}\sin\theta d\theta\int_0^{2\pi} d\phi
\int_0^{\pi}\sin\theta 'd\theta'\int_0^{2\pi} d\phi' \nonumber \\
 &  & \times\frac{\rho_a(\theta,\phi-\psi(\vec{n})-\Delta\psi(\vec{n}))\rho_a(\theta',\phi'-\psi(\vec{n}')-\Delta\psi(\vec{n}'))}
 {D(\theta,\phi,\theta',\phi';\vec{n},\vec{n}')}, \label{formulaUa}
\end{eqnarray}
with
\begin{eqnarray}
F & = & \frac{e^2}{4\pi\epsilon_0}=167100\mbox{ K \AA}, \label{refF} \label{formulaUb} \\
D(\theta,\phi,\theta',\phi';\vec{n},\vec{n}') & = & \left\{[R\cos\theta-R\cos\theta'-(n'_1-n_1)a]^2
+[R\sin\theta\cos\phi-R\sin\theta'\cos\phi'-(n'_2-n_2)b]^2 \right. \nonumber \\
 &  & \left.+[R\sin\theta\sin\phi-R\sin\theta'\sin\phi'-(n'_3-n_3)c]^2\right\}^{1/2}. \label{formulaUc}
\end{eqnarray}
\end{subequations}
\end{widetext}
The dependence on $\Delta\psi_0$ of the quantity $V^I+V^{II}$, to which ${\cal U}$ is proportional, is shown in
Fig.\ \ref{figureenergy} for both KC$_{60}$ ($Pmnn$ structure) and RbC$_{60}$ ($I2/m$ structure).  The lattice
constants used in the calculations are given in Table \ref{table1}.  The potential energy for KC$_{60}$
exhibits a double minimum, which implies that a deviation of the charge distributions from the $Pmnn$ structure
as described by the doubling mechanism of Sec.\ \ref{doubling} [Fig. \ref{fig2}(a)] is favored.  The optimal
configuration corresponds to a deviation angle of $\Delta\psi_0\approx 13^\circ$.  On the other hand, for
RbC$_{60}$, such a deviation [Fig. \ref{fig2}(b)] would never lead to an energetically
more favorable structure since the energy minimum lies at $\Delta\psi_0=0^\circ$.

\begin{table}
\caption{
Lattice constants of K- and RbC$_{60}$, taken from Ref.\ \onlinecite{Rou} ($T=6$ K values for KC$_{60}$ and $T=5$ K values for
RbC$_{60}$), and of CsC$_{60}$, taken from Ref.\ \onlinecite{Huq} ($T=20$ K values); units of {\AA}.
}
\label{table1}
\begin{ruledtabular}
\begin{tabular}{cccc}
& $a$ & $b$ & $c$ \\
\hline
KC$_{60}$ & $9.1185$ & $9.9010$ & $14.3467$ \\
RbC$_{60}$ & $9.0887$ & $10.0843$ & $14.1583$ \\
CsC$_{60}$ & $9.0968$ & $10.1895$ & $14.1351$ \\
\end{tabular}
\end{ruledtabular}
\end{table}

For completeness, the result for CsC$_{60}$ is also shown in Fig.\ \ref{figureenergy}.  Since the $I2/m$ space group angle is the same for
both Rb- and CsC$_{60}$, $\psi_{\text{RbC}_{60}}=\psi_{\text{CsC}_{60}}=46^\circ$ \cite{Rou,Huq}, the only difference between RbC$_{60}$ and
CsC$_{60}$ arises from the different lattice constants (Table \ref{table1}).  Intuitively, one would therefore expect no qualitative
difference between Rb- and CsC$_{60}$.  Indeed, Fig.\ \ref{figureenergy} confirms the similarity between Rb- and CsC$_{60}$.  Hence our model
implies the absence of a $(\vec{a}+\vec{c},\vec{b},\vec{a}-\vec{c})$ superstructure (and a concomitant metal-insulator transition) for
CsC$_{60}$.

At this point, we note that the essential dependence of the potential energy ${\cal U}$ due to
C$_{60}^-$--C$_{60}^-$ interactions on the distortion angle $\Delta\psi_0$ is introduced in Eq.\
(\ref{Deltapsi0ref}).  There, the deviation angle $\psi(\vec{n})$ at lattice site $\vec{n}$ is defined in such
a way that the experimentally found doubling scheme is automatically recovered.  However, by assigning to each
lattice site $\vec{n}$ an order parameter $S(\vec{n})=\pm 1$, corresponding to a deviation angle
$\Delta\psi(\vec{n})=\pm\Delta\psi_0$, one can show rigorously that this doubling scheme has the lowest
potential energy.  This analysis is carried out in detail in Appendix \ref{appendixA}.  There it is shown that
the average order parameter $\left<S(\vec{n})\right>$ is given by
\begin{equation}
\left<S(\vec{n})\right>=\eta(-1)^{n_1+n_3}, \label{condscheme}
\end{equation}
where $\eta$ is the order parameter amplitude.  The condensation scheme (\ref{condscheme}) is indeed equivalent
to Eq.\ (\ref{Deltapsi0ref}), expressing periodicity doubling along $\vec{a}$ and $\vec{c}$ and no change in
periodicity along $\vec{b}$.

\section{Displacements of the alkali-metal ions}
It is clear from Fig.\ \ref{figureenergy} that the valence electronic density deviations $\Delta\psi(\vec{n})$
from the $Pmnn$ structure are already sufficient to account for the periodicity doubling along the $\vec{a}$
and $\vec{c}$ axes in KC$_{60}$ observed by the x-ray scattering measurements described in Ref.\ \onlinecite{Cou}.
In this section we investigate the role of the alkali-metal ions.

The ability of the charge distributions on the C$_{60}^-$ monomers to rotate creates a picture reminiscent of
rotating molecules in molecular crystals.  It is well known that orientational motion of molecular ions in
molecular crystals influences the translational movements of neighboring counterions.  Here, we have an
analogous situation and in KC$_{60}$ one expects average center-of-mass displacements of the K$^+$ ions
induced by the angular deviations of the charge on the C$_{60}^-$ ions.

\begin{figure*}
\includegraphics{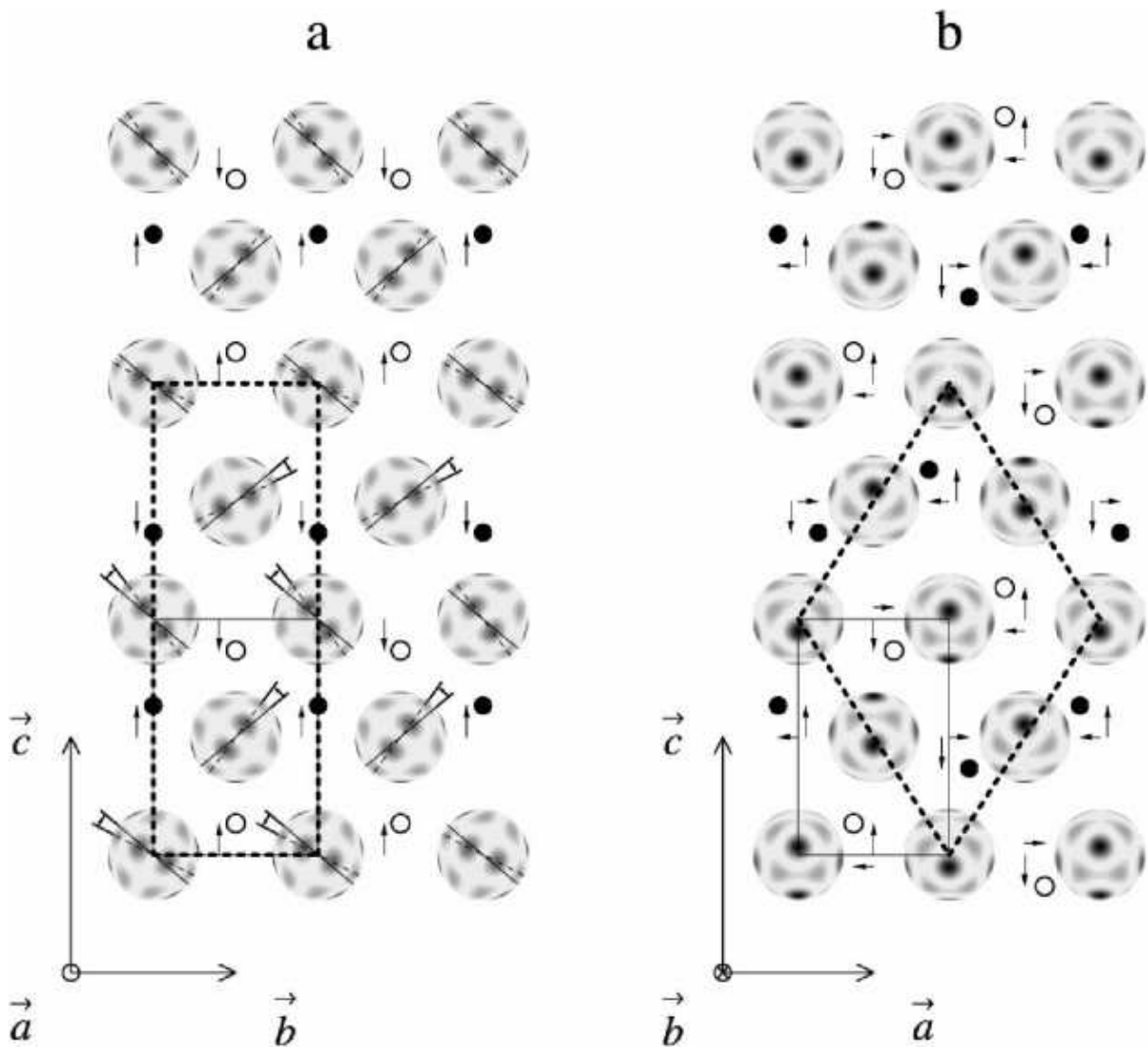} 
\caption{
The full doubling mechanism in KC$_{60}$: the charge distributions are rotated $\pm 13^\circ$ away from their
original $Pmnn$ structure, and the equilibrium positions of the alkali-metal ions are shifted, resulting in a
$(\vec{a}+\vec{c},\vec{b},\vec{a}-\vec{c})$ superstructure.  The charge distributions are represented by their
contourplot projections of Fig.\ \ref{figurechargedistribution}.  The radius of the C$_{60}^-$ units has been
reduced for clarity.  The alkali-metal ions are represented by filled and empty circles.  (a) Projection onto the
crystallographic $(\vec{b},\vec{c})$ plane (compare with Fig.\ \ref{fig2}).  (b) Projection
onto the crystallographic $(\vec{a},\vec{c})$ plane.
} 
\label{fig5} 
\end{figure*}

The theory of bilinear translation-rotation (TR) coupling in molecular crystals \cite{Mic3,Lyn} is the tool to
examine the effect of molecular orientations on lattice displacements of counterions and can be applied here as
well.  The interactions to be considered are the Coulomb attractions between the C$_{60}^-$ monomers and the
K$^+$ ions.  Full details are given in Appendix \ref{appendixB}.  The main result, Eq.\ (\ref{dispscheme}), is
that average displacements $\left<\vec{u}_{\text A}(\vec{n}_{\text A})\right>$ of the alkali-metal ions, labeled by
lattice indices $\vec{n}_{\text A}=(n_{{\text A}1},n_{{\text A}2},n_{{\text A}3})$, are found to
occur:
\begin{equation}
\left<\vec{u}_{\text A}(\vec{n}_{\text A})\right>=\left(
\begin{array}{c}
u_1 \\
0 \\
u_3
\end{array}
\right)(-1)^{k_{\text A}}. \label{mainresult}
\end{equation}
Here $k_{\text A}$ is the integer part of $n_{{\text A}1}+n_{{\text A}3}$:
\begin{equation}
n_{{\text A}1}+n_{{\text A}3}=k_{\text A}+\frac{1}{2}.
\end{equation}
Equation (\ref{mainresult}) shows that there are no displacements of the K$^+$ ions (and therefore no periodicity
doubling) along the $\vec{b}$ axis, and that along the $\vec{a}$ and $\vec{c}$ axes displacements of the K$^+$
ions happen in such a way that the periodicity doubling scheme of the charge distributions on the C$_{60}^-$
monomers is respected.  The average K$^+$ displacements---a result of the rotational deviations of the
charge distributions---constitute therefore a secondary doubling mechanism and form a part of the structural
change in polymerized KC$_{60}$.  As mentioned in the Introduction, displacements of the K$^+$ ions were
suggested by Coulon {\it et al.}\ \cite{Cou} to explain (partly) the structural phase transition.  The total doubling
mechanism, now including the K$^+$ displacements, is visualized in Fig.\ \ref{fig5}.

In Rb- and CsC$_{60}$, the charge distributions of the C$_{60}^-$ monomers do not rotate away from the original
$I2/m$ structure.  The alkali-metal ions will therefore exhibit no average displacements, keeping the $I2/m$
structure intact.

\section{Discussion and Conclusions}
In the preceding sections, we have suggested and examined the possibility of having small rotations of the valence
charge distributions on all C$_{60}^-$ monomers in the AC$_{60}$ alkali-metal fullerides.  In KC$_{60}$, these
electronic density distortions lower the potential energy and result in average center-of-mass displacements of
the K$^+$ ions.  Both the orientational deviations of the charge distributions and the translations of the
alkali-metals
occur in such a way that the ``new" structure has a doubled periodicity along the $\vec{a}$ and $\vec{c}$
directions in comparison with the ``old" structure, while the periodicity along the $\vec{b}$ direction does not
change.  These two structural changes can therefore account for the experimentally observed
$(\vec{a}+\vec{c},\vec{b},\vec{a}-\vec{c})$ superstructure in KC$_{60}$.  In RbC$_{60}$, the potential energy is
not lowered by deviations of the orientations of the charge densities.  Hence, the $I2/m$ structure will be
preserved.  (Since the charge densities do not deviate from their $I2/m$ equilibrium orientations, there is no
driving force to displace the Rb$^+$ ions, which will therefore remain at their equilibrium positions.)  To
summarize, the model we present forms a possible mechanism to explain both the experimentally observed periodicity
doubling in KC$_{60}$ and the absence of a similar doubling scheme in RbC$_{60}$.  It establishes a theoretical
basis of ``an appealing hypothesis" discussed in Ref.\ \onlinecite{Cou}, where a combination of a charge density
wave (CDW) with large correlated K displacements was suggested as a mechanism for the superstructure.

We recall that our model allows rotations of the charge distributions associated with the C$_{60}^-$ monomers in
the lattice, while the C cores remain at fixed positions, i.e., the structure formed by the C cores does not
deviate from the original $Pmnn$ or $I2/m$ structure.  This immediately launches the question why the C core
network and the electronic distribution around it would behave so ``independently" (in KC$_{60}$).  First,
we note that the angular deviation is relatively small: the energy minimum occurs at
$\Delta\psi_0\approx13^\circ$.  Furthermore, one can argue that any deviation angle $\Delta\psi_0$ different from
zero already causes an energy lowering, and that it may even be so that the cores of the C$_{60}$ clusters do
``follow" the electronic density distortions, thereby causing some restoring forces that prevent the structure
from going as far as the $\Delta\psi_0\approx13^\circ$ configuration but rather causing an equilibrium situation
at a smaller deviation angle.  A smaller equilibrium angle can also be the result of restoring forces that act
against the change in chemical bonding between C$_{60}^-$ monomers, since the chemical bonding is affected by
rotations of the electron distributions on the C$_{60}^-$ monomers.

Concerning the accompanying metal-insulator transition in KC$_{60}$ \cite{Cou}, we note that the periodicity
doubling along the $\vec{a}$ axis via the mechanism described in Sec.\ \ref{doubling} will affect the electron
transport properties along the polymer chains.  We point out that the suggested doubling scheme along the
polymerization axis can be seen as a CDW, however, not in the usual sense of a charge {\em quantity} modulation,
but a modulation of the {\em orientation} of the charge distribution along the polymer chain, which we call an
orientational charge density wave (OCDW).  The charge of every C$_{60}$ unit remains the same.  It is well known
that a one-dimensional electron gas, coupled to the underlying crystal lattice through electron-phonon
interactions, is unstable.  The Peierls instability leads to a CDW accompanied by periodic lattice distortions
\cite{Pei}.  In the case of a complete charge transfer of one electron to each anion, a half-filled band leads to
an instability with wave vector $q=2k_F=2(\pi/2a)$, which corresponds to a doubling of the unit cell from lattice
constant $a$ to $2a$ in real space.  The insulating state results from the opening of an energy gap that
separates the filled lower electron band from the empty conduction band.  In the present case of an OCDW, the
modulation of the orientations of the charge distributions along the polymer chain and the concomitant
displacements of the K$^+$ ions play the role of the lattice distortions, and the metal-insulator transition is a
consequence of the structural transformation.  It is not necessarily accompanied by displacements of the
C$_{60}^-$ monomers along the polymerization direction.  Another consequence of the rotations of the electron
densities on the C$_{60}$ units is a decrease of the transfer integrals between neighboring molecules in a polymer
chain, which also results in a reduction of the conductivity during the phase transition.

Theoretical work \cite{Mic,Ver} on the unpolymerized $\rightarrow$ polymerized phase transition in the
AC$_{60}$, A=K, Rb, Cs, compounds has revealed that the structural difference of the polymer phases (space group
$Pmnn$ for KC$_{60}$, $I2/m$ for Rb- and CsC$_{60}$) is due to the electronic quadrupolarizability of the
alkali-metals,
and not due to some other alkali-metal-specific parameters such as lattice constants or interaction strengths.  In our
view, the alkali-atom-specificity of the structural phase transition (which is present for K- but absent for
RbC$_{60}$) studied here is again not due to lattice constants, but is a direct consequence of the different space
groups of the two compounds studied.  One can therefore say that it is again the electronic quadrupolarizability
causing---indirectly---the structural difference of K- and RbC$_{60}$, since it is responsible for the
different space groups.

In conclusion, we have presented a model that (i) explains the occurrence of a second structural phase transition
in KC$_{60}$ and the absence of such a transition in RbC$_{60}$, both observed experimentally, and (ii) is a
starting point for investigations concerning the electronic properties of the alkali-metal fullerides in general and
the experimentally observed metal-insulator transition in KC$_{60}$ in particular.

\begin{acknowledgments}
We acknowledge useful discussions with P. Launois, R. Moret, and A. P\'{e}nicaud.  This work has been financially supported by the Fonds voor
Wetenschappelijk Onderzoek, Vlaanderen, and by the Bijzonder Onderzoeksfonds, Universiteit Antwerpen, UIA.
\end{acknowledgments}

\appendix
\section{Condensation Scheme for the Structural Phase Transition}\label{appendixA}
From Fig.\ \ref{figureenergy}, it follows that in KC$_{60}$, simultaneous deviations of the charge
distributions on all C$_{60}^-$ monomers, described by the doubling mechanism of Sec.\ \ref{doubling}, lead
to an energetically more favorable structure.  In this appendix, we show rigorously that this scheme will
indeed occur, without making the a priori assumption of alternating electronic deviations along the $\vec{a}$
and $\vec{c}$ axes and equal deviations along the $\vec{b}$ axis, expressed mathematically by
Eq.\ (\ref{Deltapsi0ref}).

To each site $\vec{n}$, we assign a quantity $S(\vec{n})$, which takes on the values $+1$ and $-1$,
corresponding, respectively, to a deviation angle of the electronic density $\Delta\psi(\vec{n})=+\Delta\psi_0$
and $-\Delta\psi_0$.
Allowing only the two deviation angles resulting in minima in the potential energy curve (Fig.\
\ref{figureenergy}), $|\Delta\psi_0|=13.0208^\circ$, the interaction energy $U(\vec{n},\vec{n}+\vec{\mu})$
[Eq.\ (\ref{Unnprime})] can be written as
\begin{widetext}
\begin{eqnarray}
U(\vec{n},\vec{n}+\vec{\mu}) & = &
\frac{1+S(\vec{n})}{2}U^{++}(\vec{n},\vec{n}+\vec{\mu})\frac{1+S(\vec{n}+\vec{\mu})}{2}
+\frac{1-S(\vec{n})}{2}U^{-+}(\vec{n},\vec{n}+\vec{\mu})\frac{1+S(\vec{n}+\vec{\mu})}{2} \nonumber \\
 & & +\frac{1+S(\vec{n})}{2}U^{+-}(\vec{n},\vec{n}+\vec{\mu})\frac{1-S(\vec{n}+\vec{\mu})}{2}
 +\frac{1-S(\vec{n})}{2}U^{--}(\vec{n},\vec{n}+\vec{\mu})\frac{1-S(\vec{n}+\vec{\mu})}{2},
\end{eqnarray}
\end{widetext}
where $U^{\pm+}(\vec{n},\vec{n}+\vec{\mu})$ is the value of $U(\vec{n},\vec{n}+\vec{\mu})$ when
$S(\vec{n})=\pm 1$ and $S(\vec{n}+\vec{\mu})=1$.  Analogously, $U^{\pm-}(\vec{n},\vec{n}+\vec{\mu})$ is the value
of $U(\vec{n},\vec{n}+\vec{\mu})$ when $S(\vec{n})=\pm 1$ and $S(\vec{n}+\vec{\mu})=-1$.  As in
Sec.\ \ref{potentialenergy}, we consider 14 nearest neighbors.  The energies $U^{\pm+}(\vec{n},\vec{n}+\vec{\mu})$
and $U^{\pm-}(\vec{n},\vec{n}+\vec{\mu})$ have been calculated using Eqs.\ (\ref{formulaUa})--(\ref{formulaUc})
and are listed in Table \ref{tableA1}.  Since the corner points ($\vec{n}\in\mathbb{Z}^3$) and the center point
[$\vec{n}\in(\mathbb{Z}+\frac{1}{2})^3$] of the orthorhombic cells have a different chain angle
($+\psi_{\text{KC}_{60}}$ and $-\psi_{\text{KC}_{60}}$, respectively; see Eq.\ (\ref{psiKC60}) and
Fig.\ \ref{figurestructures}), they have to be considered separately.

\begin{table*}
\caption{
Energies $U^{++}$, $U^{-+}$, $U^{+-}$ and $U^{--}$, calculated for the 14 nearest neighbors whose relative
lattice indices $\vec{\mu}$ are listed in the first column.  A distinction between the corner points (upper part of the
table) and the center point (lower part of the table) of the orthorhombic cells has to be made.  Because of
symmetry reasons, only a limited number of different numerical values occurs.
}
\label{tableA1}
\begin{ruledtabular}
\begin{tabular}{lllll}
\multicolumn{5}{c}{
$\vec{n}\in\mathbb{Z}^3$
} \\
$\vec{\mu}$ & $U^{++}(\vec{n},\vec{n}+\vec{\mu})$ & $U^{-+}(\vec{n},\vec{n}+\vec{\mu})$ & $U^{+-}(\vec{n},\vec{n}+\vec{\mu})$ & $U^{--}(\vec{n},\vec{n}+\vec{\mu})$ \\
\hline
$(1,0,0)$ & $J_a^+=18203$ K & $J_a^-=18200$ K & $J_a^-$ & $J_a^+$ \\
$(0,1,0)$ & $J_b^{++}=16946$ K & $J_b^-=17060$ K & $J_b^-$ & $J_b^{--}=17160$ K \\
$(-1,0,0)$ & $J_a^+$ & $J_a^-$ & $J_a^-$ & $J_a^+$ \\
$(0,-1,0)$ & $J_b^{++}$ & $J_b^-$ & $J_b^-$ & $J_b^{--}$ \\
$(0,0,1)$ & $J_c^{++}=11708$ K & $J_c^-=11669$ K & $J_c^-$ & $J_c^{--}=11631$ K \\
$(0,0,-1)$ & $J_c^{++}$ & $J_c^-$ & $J_c^-$ & $J_c^{--}$ \\
$(\frac{1}{2},\frac{1}{2},\frac{1}{2})$ & $J_{abc}^+=16980$ K & $J_{abc}^{-+}=16954$ K & $J_{abc}^{+-}=17003$ K & $J_{abc}^+$ \\
$(-\frac{1}{2},\frac{1}{2},\frac{1}{2})$ & $J_{abc}^+$ & $J_{abc}^{-+}$ & $J_{abc}^{+-}$ & $J_{abc}^+$ \\
$(-\frac{1}{2},-\frac{1}{2},\frac{1}{2})$ & $J_{abc}^+$ & $J_{abc}^{-+}$ & $J_{abc}^{+-}$ & $J_{abc}^+$ \\
$(\frac{1}{2},-\frac{1}{2},\frac{1}{2})$ & $J_{abc}^+$ & $J_{abc}^{-+}$ & $J_{abc}^{+-}$ & $J_{abc}^+$ \\
$(\frac{1}{2},\frac{1}{2},-\frac{1}{2})$ & $J_{abc}^+$ & $J_{abc}^{-+}$ & $J_{abc}^{+-}$ & $J_{abc}^+$ \\
$(-\frac{1}{2},\frac{1}{2},-\frac{1}{2})$ & $J_{abc}^+$ & $J_{abc}^{-+}$ & $J_{abc}^{+-}$ & $J_{abc}^+$ \\
$(-\frac{1}{2},-\frac{1}{2},-\frac{1}{2})$ & $J_{abc}^+$ & $J_{abc}^{-+}$ & $J_{abc}^{+-}$ & $J_{abc}^+$ \\
$(\frac{1}{2},-\frac{1}{2},-\frac{1}{2})$ & $J_{abc}^+$ & $J_{abc}^{-+}$ & $J_{abc}^{+-}$ & $J_{abc}^+$ \\
\multicolumn{5}{c}{
$\vec{n}\in(\mathbb{Z}+\frac{1}{2})^3$
} \\
$\vec{\mu}$ & $U^{++}(\vec{n},\vec{n}+\vec{\mu})$ & $U^{-+}(\vec{n},\vec{n}+\vec{\mu})$ & $U^{+-}(\vec{n},\vec{n}+\vec{\mu})$ & $U^{--}(\vec{n},\vec{n}+\vec{\mu})$ \\
\hline
$(1,0,0)$ & $J_a^+$ & $J_a^-$ & $J_a^-$ & $J_a^+$ \\
$(0,1,0)$ & $J_b^{--}$ & $J_b^-$ & $J_b^-$ & $J_b^{++}$ \\
$(-1,0,0)$ & $J_a^+$ & $J_a^-$ & $J_a^-$ & $J_a^+$ \\
$(0,-1,0)$ & $J_b^{--}$ & $J_b^-$ & $J_b^-$ & $J_b^{++}$ \\
$(0,0,1)$ & $J_c^{--}$ & $J_c^-$ & $J_c^-$ & $J_c^{++}$ \\
$(0,0,-1)$ & $J_c^{--}$ & $J_c^-$ & $J_c^-$ & $J_c^{++}$ \\
$(\frac{1}{2},\frac{1}{2},\frac{1}{2})$ & $J_{abc}^+$ & $J_{abc}^{+-}$ & $J_{abc}^{-+}$ & $J_{abc}^+$ \\
$(-\frac{1}{2},\frac{1}{2},\frac{1}{2})$ & $J_{abc}^+$ & $J_{abc}^{+-}$ & $J_{abc}^{-+}$ & $J_{abc}^+$ \\
$(-\frac{1}{2},-\frac{1}{2},\frac{1}{2})$ & $J_{abc}^+$ & $J_{abc}^{+-}$ & $J_{abc}^{-+}$ & $J_{abc}^+$ \\
$(\frac{1}{2},-\frac{1}{2},\frac{1}{2})$ & $J_{abc}^+$ & $J_{abc}^{+-}$ & $J_{abc}^{-+}$ & $J_{abc}^+$ \\
$(\frac{1}{2},\frac{1}{2},-\frac{1}{2})$ & $J_{abc}^+$ & $J_{abc}^{+-}$ & $J_{abc}^{-+}$ & $J_{abc}^+$ \\
$(-\frac{1}{2},\frac{1}{2},-\frac{1}{2})$ & $J_{abc}^+$ & $J_{abc}^{+-}$ & $J_{abc}^{-+}$ & $J_{abc}^+$ \\
$(-\frac{1}{2},-\frac{1}{2},-\frac{1}{2})$ & $J_{abc}^+$ & $J_{abc}^{+-}$ & $J_{abc}^{-+}$ & $J_{abc}^+$ \\
$(\frac{1}{2},-\frac{1}{2},-\frac{1}{2})$ & $J_{abc}^+$ & $J_{abc}^{+-}$ & $J_{abc}^{-+}$ & $J_{abc}^+$
\end{tabular}
\end{ruledtabular}
\end{table*}

The total energy ${\cal U}$ [Eq.\ (\ref{U})] is then obtained by summing $U(\vec{n},\vec{n}+\vec{\mu})$ over
the whole lattice.  It is convenient to write the result in Fourier space.  Defining the discrete Fourier
transform of $S(\vec{n})$ by
\begin{subequations}
\begin{eqnarray}
S(\vec{n}) & = & \frac{1}{\sqrt{{\cal N}}}\sum_{\vec{q}}e^{i\vec{q}\cdot\vec{X}(\vec{n})}S(\vec{q}),
\label{FourierSn} \\
S(\vec{q}) & = & \frac{1}{\sqrt{{\cal N}}}\sum_{\vec{n}}e^{-i\vec{q}\cdot\vec{X}(\vec{n})}S(\vec{n}),
\label{FourierSq}
\end{eqnarray}
\end{subequations}
we obtain
\begin{eqnarray}
{\cal U} & = & \frac{1}{2}\sum_{\vec{n}}\sum_{\vec{\mu}}U(\vec{n},\vec{n}+\vec{\mu}) \nonumber \\
& = & \frac{1}{2}\sum_{\vec{q}}J(\vec{q})S(\vec{q})S(-\vec{q})+C \label{JqandC}.
\end{eqnarray}
Here we have split the summation over $\vec{n}$ into two parts: corner points $\vec{n}\in\mathbb{Z}^3$ and
center points $\vec{n}\in(\mathbb{Z}+\frac{1}{2})^3$.  The summation over $\vec{n}$ in Eq.\ (\ref{FourierSq})
is understood to be a summation over corner points---of which there are ${\cal N}=N/2$---only.  In
Eq.\ (\ref{JqandC}), $C$ is an irrelevant constant and
\begin{subequations}
\begin{eqnarray}
J(\vec{q}) & = & J_a\cos(q_Xa)+J_b\cos(q_Yb)+J_c\cos(q_Zc) \nonumber \\
 & & +J_{abc}\cos(q_X\frac{a}{2})\cos(q_Y\frac{b}{2})\cos(q_Z\frac{c}{2}), \\
J_a & = & 2(J_a^+-J_a^-), \\
J_b & = & J_b^{++}-2J_b^-+J_b^{--}, \\
J_c & = & J_c^{++}-2J_c^-+J_c^{--}, \\
J_{abc} & = & 4(2J_{abc}^+-J_{abc}^{-+}-J_{abc}^{+-}).
\end{eqnarray}
\end{subequations}
The coefficients $J_a^+,\ldots,J_{abc}^{+-}$ are related to the potential energies $U^{++},\ldots,U^{--}$ as is
indicated in Table \ref{tableA1}.  We now determine the absolute minimum of the function $J(\vec{q})$ in
reciprocal space.
Taking into account the numerical values of $J_a=6$ K, $J_b=-14$ K,
$J_c=1$ K, and $J_{abc}=12$ K, one finds that the absolute minimum lies at
$\vec{q}=(q_X,q_Y,q_Z)=(\frac{\pi}{a},0,\frac{\pi}{c})\equiv\vec{q_B}$.
The dominance of $J(\vec{q}_B)$ leads to a condensation of $S(\vec{q})$ at $\vec{q}=\vec{q}_B$:
\begin{equation}
\left<S(\vec{q})\right>=\eta\sqrt{{\cal N}}\delta_{\vec{q},\vec{q}_B}. \label{condensation}
\end{equation}
Here $\eta$ is the order parameter amplitude.  The condensation scheme in Fourier space (\ref{condensation})
corresponds to the following real space condensation scheme:
\begin{equation}
\left<S(\vec{n})\right>=\eta\cos[\vec{q}_B\cdot\vec{X}(\vec{n})]=\eta(-1)^{n_1+n_3}. \label{condensationrealspace}
\end{equation}

\section{Translation-rotation coupling in KC$_{60}$}\label{appendixB}
In this appendix, we examine the coupling between the orientational deviations of the charge distributions on
the C$_{60}^-$ monomers---lowering the crystal's potential energy in KC$_{60}$ (see Fig.\
\ref{figureenergy})---and displacements of the
K$^{+}$ ions.  We use concepts of the theory of bilinear translation-rotation (TR) coupling in molecular
crystals \cite{Mic3,Lyn}, which is generally used to determine the influence of molecular orientations on lattice
displacements of counterions.

The starting point is the potential energy of a C$_{60}^-$ monomer and a K$^+$ ion:
\begin{eqnarray}
U(\vec{n},\vec{n}_{\text A}) & = & \frac{1}{4\pi\epsilon_0}\int d\vec{r}\int d\vec{r}_{\text A} \nonumber \\
 & \times & \frac{\rho(\vec{r};\psi(\vec{n})+\Delta\psi(\vec{n}))\rho_{\text A}(\vec{r}_{\text A})}
{|\vec{r}-\vec{r}_{\text A}-\vec{X}(\vec{n}_{\text A}-\vec{n})-\vec{u}_{\text A}(\vec{n}_{\text A})|}, \label{UnnA}
\end{eqnarray}
where the subscript $_{\text A}$ refers to the alkali-metal ion.  Similarly as in Eq.\ (\ref{Unnprime}), the integration
variables $\vec{r}$ and $\vec{r}_{\text A}$ in Eq.\ (\ref{UnnA}) refer to the {\em local} coordinate systems associated
with the lattice sites $\vec{n}$ and $\vec{n}_{\text A}$, respectively, being the reason for the appearance of the
relative lattice vector $\vec{X}(\vec{n}_{\text A}-\vec{n})$ and the alkali-metal lattice displacement vector
$\vec{u}_{\text A}(\vec{n}_{\text A})$.
To be consistent with the earlier convention
\begin{equation}
\left\{\begin{array}{l}
n_1\in\mathbb{Z} \\
n_2\in\mathbb{Z} \\
n_3\in\mathbb{Z}
\end{array}\right\}
\mbox{ or }
\left\{\begin{array}{l}
n_1\in\mathbb{Z}+\frac{1}{2} \\
n_2\in\mathbb{Z}+\frac{1}{2} \\
n_3\in\mathbb{Z}+\frac{1}{2}
\end{array}\right\},
\end{equation}
one must have for the alkali-metal lattice indices $\vec{n}_{\text A}$:
\begin{equation}
\left\{\begin{array}{l}
n_{{\text A}1}\in\mathbb{Z} \\
n_{{\text A}2}\in\mathbb{Z} \\
n_{{\text A}3}\in\mathbb{Z}+\frac{1}{2}
\end{array}\right\}
\mbox{ or }
\left\{\begin{array}{l}
n_{{\text A}1}\in\mathbb{Z}+\frac{1}{2} \\
n_{{\text A}2}\in\mathbb{Z}+\frac{1}{2} \\
n_{{\text A}3}\in\mathbb{Z}
\end{array}\right\}. \label{alkalilattice}
\end{equation}

We treat the K$^+$ ion as a point charge and write for its charge distribution:
\begin{equation}
\rho_{\text A}(\vec{r}_{\text A})=e\delta(\vec{r}_{\text A}). \label{rhoA}
\end{equation}
Working with the charge distribution
$\rho(\vec{r};\psi(\vec{n})+\Delta\psi(\vec{n}))=\rho(r,\theta,\phi-\psi(\vec{n})-\Delta\psi(\vec{n}))$ of
Sec.\ \ref{charge} for the C$_{60}^-$ monomer and taking into account Eq.\ (\ref{rhoA}), we get the following
expression for $U(\vec{n},\vec{n}_{\text A})$:
\begin{widetext}
\begin{subequations}
\begin{eqnarray}
U(\vec{n},\vec{n}_{\text A}) & = & F\int_0^\pi\sin\theta
d\theta\int_0^{2\pi}d\phi\frac{\rho_a(\theta,\phi-\psi(\vec{n})-\Delta\psi(\vec{n}))}
{d(\theta,\phi;\vec{n},\vec{n}_{\text A})}, \\
d(\theta,\phi;\vec{n},\vec{n}_{\text A}) & = & \left\{[R\cos\theta-(n_{{\text A}1}-n_1)a-u_{{\text A}1}(\vec{n}_{\text A})]^2+[R\sin\theta\cos\phi-(n_{{\text A}2}-n_2)b-u_{{\text A}2}(\vec{n}_{\text A})]^2\right. \nonumber \\
 & & \left.+[R\sin\theta\sin\phi-(n_{{\text A}3}-n_3)c-u_{{\text A}3}(\vec{n}_{\text A})]^2
\right\}^{1/2},
\end{eqnarray}
\end{subequations}
\end{widetext}
where the constant $F$ is given by Eq.\ (\ref{refF}).

We consider small center-of-mass displacements of the K$^+$ ions and expand $U(\vec{n},\vec{n}_{\text A})$ in terms of
the components of $\vec{u}_{\text A}(\vec{n}_{\text A})$, retaining only the zeroth- and first-order terms:
\begin{eqnarray}
U(\vec{n},\vec{n}_{\text A}) & = & \left.U(\vec{n},\vec{n}_{\text A})\right|_{\vec{u}_{\text A}(\vec{n}_{\text
A})=\vec{0}} \nonumber \\
 & + & \sum_{i=1}^3\left.\frac{\partial U(\vec{n},\vec{n}_{\text A})}{\partial
u_{{\text A}i}(\vec{n}_{\text A})}\right|_{\vec{u}_{\text A}(\vec{n}_{\text A})=\vec{0}}u_{{\text A}i}(\vec{n}_{\text A}). \label{expansion}
\end{eqnarray}

As in Appendix \ref{appendixA}, we introduce the quantity $S(\vec{n})=\pm 1$, corresponding to
$\Delta\psi(\vec{n})=\pm|\Delta\psi_0|$, with $|\Delta\psi_0|=13.0208^\circ$.  The interaction energy
$U(\vec{n},\vec{n}_{\text A})$ can then be written as
\begin{equation}
U(\vec{n},\vec{n}_{\text A})=\frac{1+S(\vec{n})}{2}U^+(\vec{n},\vec{n}_{\text A})
+\frac{1-S(\vec{n})}{2}U^-(\vec{n},\vec{n}_{\text A}),
\end{equation}
where $U^\pm(\vec{n},\vec{n}_{\text A})$ is the value of $U(\vec{n},\vec{n}_{\text A})$ when
$S(\vec{n})=\pm 1$.  The expansion (\ref{expansion}) becomes
\begin{subequations}
\begin{widetext}
\begin{eqnarray}
U(\vec{n},\vec{n}_{\text A}) & = & \frac{1+S(\vec{n})}{2}\left\{V^+(\vec{n},\vec{n}_{\text A})
+\sum_{i=1}^3v_i^+(\vec{n},\vec{n}_{\text A})u_{{\text A}i}(\vec{n}_{\text
A})\right\} \nonumber \\
 & & +\frac{1-S(\vec{n})}{2}\left\{V^-(\vec{n},\vec{n}_{\text A})
+\sum_{i=1}^3v_i^-(\vec{n},\vec{n}_{\text A})u_{{\text A}i}(\vec{n}_{\text A})\right\},
\end{eqnarray}
\end{widetext}
with
\begin{eqnarray}
V^\pm(\vec{n},\vec{n}_{\text A}) & = & \left.U^\pm(\vec{n},\vec{n}_{\text A})\right|_{\vec{u}_{\text A}(\vec{n}_{\text A})=\vec{0}}, \\
v_i^\pm(\vec{n},\vec{n}_{\text A}) & = & \left.\frac{\partial U^\pm(\vec{n},\vec{n}_{\text A})}{\partial
u_{{\text A}i}(\vec{n}_{\text A})}\right|_{\vec{u}_{\text A}(\vec{n}_{\text A})=\vec{0}}.
\end{eqnarray}
\end{subequations}
The contribution ${\cal U}_{\text{C$_{60}$-A}}$ of all electrostatic C$_{60}^-$-K$^+$ interactions to the
potential energy is obtained by summing $U(\vec{n},\vec{n}_{\text A})$ over the whole crystal lattice:
\begin{subequations}
\begin{equation}
{\cal U}_{\text{C$_{60}$-A}}={\cal U}^{\text R}+{\cal U}^{\text{TR}},
\end{equation}
where
\begin{widetext}
\begin{eqnarray}
{\cal U}^{\text R} & = & \sum_{\vec{n}}\sum_{\vec{\mu}_{\text A}}\left\{
\frac{1+S(\vec{n})}{2}V^+(\vec{n},\vec{n}+\vec{\mu}_{\text A})+\frac{1-S(\vec{n})}{2}V^-(\vec{n},\vec{n}+\vec{\mu}_{\text A})
\right\}, \\
{\cal U}^{\text{TR}} & = & \sum_{\vec{n}}\sum_{\vec{\mu}_{\text A}}\left\{
\frac{1+S(\vec{n})}{2}\sum_{i=1}^3v_i^+(\vec{n},\vec{n}+\vec{\mu}_{\text A})u_{{\text
A}i}(\vec{n}+\vec{\mu}_{\text A})+\frac{1-S(\vec{n})}{2}\sum_{i=1}^3v_i^-(\vec{n},\vec{n}+\vec{\mu}_{\text A})u_{{\text A}i}(\vec{n}+\vec{\mu}_{\text A})
\right\}.
\end{eqnarray}
\end{widetext}
\end{subequations}

For a given C$_{60}^-$ monomer, we limit ourselves to the six nearest alkali-metal neighbors.  The values of
$V^\pm(\vec{n},\vec{n}+\vec{\mu}_{\text A})$ and $v_i^\pm(\vec{n},\vec{n}+\vec{\mu}_{\text A})$, $i=1,2,3$ are
listed in Table \ref{tableB1}.  As in Appendix \ref{appendixA}, a distinction has to be made between the corner
points ($\vec{n}\in\mathbb{Z}^3$) and the center point [$\vec{n}\in(\mathbb{Z}+\frac{1}{2})^3$] of the
orthorhombic cells.

\begin{table*}
\caption{
Energies $V^\pm$ and derivatives of energies $v_i^\pm$, $i=1,2,3$, calculated for the six nearest alkali-metal
neighbors with relative lattice indices $\vec{\mu}_{\text A}$.  A distinction between
the corner points (upper part of the table) and the center point (lower part of the table) of the orthorhombic
cells has to be made.  Only a limited number of different numerical values occurs.
}
\label{tableB1}
\begin{ruledtabular}
\begin{tabular}{lllll}
\multicolumn{5}{c}{
$\vec{n}\in\mathbb{Z}^3$
} \\
$\vec{\mu}_{\text A}$ & $V^+(\vec{n},\vec{n}+\vec{\mu}_{\text A})$ & $v_1^+(\vec{n},\vec{n}+\vec{\mu}_{\text A})$ &
$v_2^+(\vec{n},\vec{n}+\vec{\mu}_{\text A})$ & $v_3^+(\vec{n},\vec{n}+\vec{\mu}_{\text A})$ \\
\hline
$(\frac{1}{2},\frac{1}{2},0)$ & $V_{ab}^+=24665.3$ K & $-v_{1ab}^+=-2481.7$ K \AA$^{-1}$ & $-v_{2ab}^+=-2606.9$ K
\AA$^{-1}$ & $+v_{3ab}^+=85.8$ K \AA$^{-1}$ \\
$(-\frac{1}{2},\frac{1}{2},0)$ & $V_{ab}^+$ & $+v_{1ab}^+$ & $-v_{2ab}^+$ & $+v_{3ab}^+$ \\
$(-\frac{1}{2},-\frac{1}{2},0)$ & $V_{ab}^+$ & $+v_{1ab}^+$ & $+v_{2ab}^+$ & $-v_{3ab}^+$ \\
$(\frac{1}{2},-\frac{1}{2},0)$ & $V_{ab}^+$ & $-v_{1ab}^+$ & $+v_{2ab}^+$ & $-v_{3ab}^+$ \\
$(0,0,\frac{1}{2})$ & $V_{c}^+=23685.6$ K & $0$ K \AA$^{-1}$ & $+v_{2c}^+=58.3$ K \AA$^{-1}$ & $-v_{3c}^+=-3476.6$ K \AA$^{-1}$ \\
$(0,0,-\frac{1}{2})$ & $V_{c}^+$ & $0$ K \AA$^{-1}$ & $-v_{2c}^+$ & $+v_{3c}^+$ \\
$\vec{\mu}_{\text A}$ & $V^-(\vec{n},\vec{n}+\vec{\mu}_{\text A})$ & $v_1^-(\vec{n},\vec{n}+\vec{\mu}_{\text A})$ &
$v_2^-(\vec{n},\vec{n}+\vec{\mu}_{\text A})$ & $v_3^-(\vec{n},\vec{n}+\vec{\mu}_{\text A})$ \\
\hline
$(\frac{1}{2},\frac{1}{2},0)$ & $V_{ab}^-=24821.3$ K & $-v_{1ab}^-=-2531.4$ K \AA$^{-1}$ & $-v_{2ab}^-=-2636.5$
K \AA$^{-1}$ & $+v_{3ab}^-=49.1$ K \AA$^{-1}$ \\
$(-\frac{1}{2},\frac{1}{2},0)$ & $V_{ab}^-$ & $+v_{1ab}^-$ & $-v_{2ab}^-$ & $+v_{3ab}^-$ \\
$(-\frac{1}{2},-\frac{1}{2},0)$ & $V_{ab}^-$ & $+v_{1ab}^-$ & $+v_{2ab}^-$ & $-v_{3ab}^-$ \\
$(\frac{1}{2},-\frac{1}{2},0)$ & $V_{ab}^-$ & $-v_{1ab}^-$ & $+v_{2ab}^-$ & $-v_{3ab}^-$ \\
$(0,0,\frac{1}{2})$ & $V_{c}^-=23279.3$ K & $0$ K \AA$^{-1}$ & $+v_{2c}^-=148.3$ K \AA$^{-1}$ &
$-v_{3c}^-=-3253.7$ K \AA$^{-1}$ \\
$(0,0,-\frac{1}{2})$ & $V_{c}^-$ & $0$ K \AA$^{-1}$ & $-v_{2c}^-$ & $+v_{3c}^-$ \\
\multicolumn{5}{c}{
$\vec{n}\in(\mathbb{Z}+\frac{1}{2})^3$
} \\
$\vec{\mu}_{\text A}$ & $V^+(\vec{n},\vec{n}+\vec{\mu}_{\text A})$ & $v_1^+(\vec{n},\vec{n}+\vec{\mu}_{\text A})$ &
$v_2^+(\vec{n},\vec{n}+\vec{\mu}_{\text A})$ & $v_3^+(\vec{n},\vec{n}+\vec{\mu}_{\text A})$ \\
\hline
$(\frac{1}{2},\frac{1}{2},0)$ & $V_{ab}^-$ & $-v_{1ab}^-$ & $-v_{2ab}^-$ & $-v_{3ab}^-$ \\
$(-\frac{1}{2},\frac{1}{2},0)$ & $V_{ab}^-$ & $+v_{1ab}^-$ & $-v_{2ab}^-$ & $-v_{3ab}^-$ \\
$(-\frac{1}{2},-\frac{1}{2},0)$ & $V_{ab}^-$ & $+v_{1ab}^-$ & $+v_{2ab}^-$ & $+v_{3ab}^-$ \\
$(\frac{1}{2},-\frac{1}{2},0)$ & $V_{ab}^-$ & $-v_{1ab}^-$ & $+v_{2ab}^-$ & $+v_{3ab}^-$ \\
$(0,0,\frac{1}{2})$ & $V_{c}^-$ & $0$ K \AA$^{-1}$ & $-v_{2c}^-$ & $-v_{3c}^-$ \\
$(0,0,-\frac{1}{2})$ & $V_{c}^-$ & $0$ K \AA$^{-1}$ & $+v_{2c}^-$ & $+v_{3c}^-$ \\
$\vec{\mu}_{\text A}$ & $V^-(\vec{n},\vec{n}+\vec{\mu}_{\text A})$ & $v_1^-(\vec{n},\vec{n}+\vec{\mu}_{\text A})$ &
$v_2^-(\vec{n},\vec{n}+\vec{\mu}_{\text A})$ & $v_3^-(\vec{n},\vec{n}+\vec{\mu}_{\text A})$ \\
\hline
$(\frac{1}{2},\frac{1}{2},0)$ & $V_{ab}^+$ & $-v_{1ab}^+$ & $-v_{2ab}^+$ & $-v_{3ab}^+$ \\
$(-\frac{1}{2},\frac{1}{2},0)$ & $V_{ab}^+$ & $+v_{1ab}^+$ & $-v_{2ab}^+$ & $-v_{3ab}^+$ \\
$(-\frac{1}{2},-\frac{1}{2},0)$ & $V_{ab}^+$ & $+v_{1ab}^+$ & $+v_{2ab}^+$ & $+v_{3ab}^+$ \\
$(\frac{1}{2},-\frac{1}{2},0)$ & $V_{ab}^+$ & $-v_{1ab}^+$ & $+v_{2ab}^+$ & $+v_{3ab}^+$ \\
$(0,0,\frac{1}{2})$ & $V_{c}^+$ & $0$ K \AA$^{-1}$ & $-v_{2c}^+$ & $-v_{3c}^+$ \\
$(0,0,-\frac{1}{2})$ & $V_{c}^+$ & $0$ K \AA$^{-1}$ & $+v_{2c}^+$ & $+v_{3c}^+$
\end{tabular}
\end{ruledtabular}
\end{table*}

Introducing the discrete Fourier transforms of the alkali-metal ion displacements,
\begin{subequations}
\begin{eqnarray}
\vec{u}_{\text A}(\vec{n_{\text A}}) & = & \frac{1}{\sqrt{{\cal N}m_{\text A}}}\sum_{\vec{q}}e^{i\vec{q}\cdot\vec{X}(\vec{n}_{\text A})}\vec{u}_{\text A}(\vec{q}), \\
\vec{u}_{\text A}(\vec{q}) & = & \sqrt{\frac{m_{\text A}}{\cal N}}\sum_{\vec{n}_{\text A}}e^{-i\vec{q}\cdot\vec{X}(\vec{n}_{\text A})}\vec{u}_{\text A}(\vec{n}_{\text A}),
\end{eqnarray}
\end{subequations}
where $m_{\text A}$ is the mass of the K$^+$ ion, and using in addition the discrete Fourier transform of
$S(\vec{n})$, defined by Eqs.\ (\ref{FourierSn}) and (\ref{FourierSq}), we get for the TR term of the potential energy
\begin{subequations}
\begin{equation}
{\cal U}^{\text{TR}}=\frac{1}{\sqrt{m_{\text A}}}\sum_{\vec{q}}S(-\vec{q})\vec{v}(\vec{q})\cdot\vec{u}_{\text A}(\vec{q}),
\end{equation}
with
\begin{widetext}
\begin{equation}
\vec{v}(\vec{q})=i\left(
\begin{array}{c}
-2{\cal D}_{1ab}\sin(q_X\frac{a}{2})\cos(q_Y\frac{b}{2})
[1-e^{i(q_X\frac{a}{2}+q_Y\frac{b}{2}+q_Z\frac{c}{2})}] \\
-2{\cal D}_{2ab}\cos(q_X\frac{a}{2})\sin(q_Y\frac{b}{2})
[1-e^{i(q_X\frac{a}{2}+q_Y\frac{b}{2}+q_Z\frac{c}{2})}]
+{\cal D}_{2c}\sin(q_Z\frac{c}{2})
[1+e^{i(q_X\frac{a}{2}+q_Y\frac{b}{2}+q_Z\frac{c}{2})}] \\
2{\cal D}_{3ab}\cos(q_X\frac{a}{2})\sin(q_Y\frac{b}{2})
[1+e^{i(q_X\frac{a}{2}+q_Y\frac{b}{2}+q_Z\frac{c}{2})}]
-{\cal D}_{3c}\sin(q_Z\frac{c}{2})
[1-e^{i(q_X\frac{a}{2}+q_Y\frac{b}{2}+q_Z\frac{c}{2})}]
\end{array}
\right)
\end{equation}
\end{widetext}
and
\begin{eqnarray}
{\cal D}_{1ab} & = & v_{1ab}^+-v_{1ab}^-, \\
{\cal D}_{2ab} & = & v_{2ab}^+-v_{2ab}^-, \\
{\cal D}_{2c} & = & v_{2c}^+-v_{2c}^-, \\
{\cal D}_{3ab} & = & v_{3ab}^+-v_{3ab}^-, \\
{\cal D}_{3c} & = & v_{3c}^+-v_{3c}^-.
\end{eqnarray}
\end{subequations}

In the theory of bilinear TR coupling in orientationally disordered crystals \cite{Lyn,Mic4}, it is shown that the
minimal potential energy for a given orientational configuration $\{S(\vec{q})\}$ is obtained when
\begin{equation}
\vec{u}_{\text A}(\vec{q})=-M^{-1}(\vec{q})\vec{v}(-\vec{q})S(\vec{q}),
\end{equation}
where $M(\vec{q})$ is the dynamical matrix of the orthorhombic crystal in absence of TR coupling.  Using for
$S(\vec{q})$ the minimal potential energy condensation scheme of Appendix \ref{appendixA},
Eq.\ (\ref{condensation}), we find for the displacements in reciprocal space:
\begin{equation}
\left<\vec{u}_{\text A}(\vec{q})\right>=-M^{-1}(\vec{q}=\vec{q}_B)i\left(
\begin{array}{c}
-4{\cal D}_{1ab}\\
0\\
-2{\cal D}_{3c}
\end{array}
\right)
\eta\sqrt{{\cal N}}\delta_{\vec{q},\vec{q}_B}. \label{schemeq}
\end{equation}
The numerical values of ${\cal D}_{1ab}$ and ${\cal D}_{3c}$ follow from Table \ref{tableB1}:
${\cal D}_{1ab}=-49.7$ K\ \AA$^{-1}$ and ${\cal D}_{3c}=222.9$ K\ \AA$^{-1}$.  Since
$\vec{q}_B\equiv(\frac{\pi}{a},0,\frac{\pi}{c})$, the scheme (\ref{schemeq}) becomes in real space
\begin{eqnarray}
\left<\vec{u}_{\text A}(\vec{n}_{\text A})\right> & = & -\frac{1}{\sqrt{m_{\text A}}}M^{-1}(\vec{q}=\vec{q}_B)\left(
\begin{array}{c}
-4{\cal D}_{1ab}\\
0\\
-2{\cal D}_{3c}
\end{array}
\right)
\eta i \nonumber \\
 & & \times e^{-i\pi(n_{{\text A}1}+n_{{\text A}3})}.
\end{eqnarray}
The alkali-metal ions are located on a sublattice obeying
$n_{{\text A}1}+n_{{\text A}3}=k_{\text A}+\frac{1}{2}$ with $k_{\text A}\in\mathbb{Z}$
[see Eq.\ (\ref{alkalilattice})].  Therefore, one has $e^{-i\pi(n_{A1}+n_{A3})}=-i(-1)^{k_{\text A}}$, yielding
\begin{eqnarray}
\left<\vec{u}_{\text A}(\vec{n}_{\text A})\right> & = & -\frac{1}{\sqrt{m_{\text A}}}M^{-1}(\vec{q}=\vec{q}_B)\left(
\begin{array}{c}
-4{\cal D}_{1ab}\\
0\\
-2{\cal D}_{3c}
\end{array}
\right)
\eta(-1)^{k_{\text A}} \nonumber \\
 & = & \left(\begin{array}{c}
 u_1 \\
 0 \\
 u_3
 \end{array}\right)(-1)^{k_{\text A}}. \label{dispscheme}
\end{eqnarray}


\begin{thebibliography}{99}
\bibitem{Win}
J. Winter and H. Kuzmany, Solid State Commun. {\bf 84}, 935 (1992); Q. Zhu, O. Zhou, J.E. Fischer,
A.R. McGhie, W.J. Romanov, R.M. Strongin, M.A. Cichy, and A.B. Smith III, Phys. Rev. B {\bf 47},
13948 (1993).

\bibitem{Pek}
S. Pekker, L. Forr\'{o}, L. Mih\'{a}ly, and A. J\'{a}nossy, Solid State Commun. {\bf 90}, 349 (1994).

\bibitem{Zhu}
Q. Zhu, D.E. Cox, and J.E. Fischer, Phys. Rev. B {\bf 51}, 3966 (1995).

\bibitem{Osz}
G. Oszl\'{a}nyi, G. Bortel, G. Faigel, M. Tegze, L. Gr\'{a}n\'{a}sy, S. Pekker, P.W. Stephens, G. Bendele, R.
Dinnebier, G. Mih\'{a}ly, A. J\'{a}nossy, O. Chauvet, and L. Forr\'{o}, Phys. Rev. B {\bf 51}, 12228 (1995).

\bibitem{Ste}
P.W. Stephens, G. Bortel, G. Faigel, M. Tegze, A. J\'{a}nossy, S. Pekker, G. Oszl\'{a}nyi, and L. Forr\'{o},
Nature (London) {\bf 370}, 636 (1994).

\bibitem{Lau1}
P. Launois, R. Moret, J. Hone, and A. Zettl, Phys. Rev. Lett. {\bf 81}, 4420 (1998).

\bibitem{Rou}
S. Rouzi\`{e}re, S. Margadonna, K. Prassides, and A.N. Fitch, Europhys. Lett. {\bf 51}, 314 (2000).

\bibitem{For}
L. Forr\'{o} and L. Mih\'{a}ly, Rep. Prog. Phys. {\bf 64}, 649 (2001).

\bibitem{Mic}
K.H. Michel and A.V. Nikolaev, Phys. Rev. Lett. {\bf 85}, 3197 (2000).

\bibitem{Ver}
B. Verberck, K.H. Michel, and A.V. Nikolaev, J. Chem. Phys. {\bf 116}, 10462 (2002).

\bibitem{Cou}
C. Coulon, A. P\'{e}nicaud, R. Cl\'{e}rac, R. Moret, P. Launois, and J. Hone, Phys. Rev. Lett. {\bf 86}, 4346 (2001).

\bibitem{Lau}
P. Launois and R. Moret (private communication).

\bibitem{Huq}
A. Huq, P.W. Stephens, G.M. Bendele, and R.M. Ibberson, Chem. Phys. Lett. {\bf 347}, 13 (2001).

\bibitem{Had}
R.C. Haddon, L.E. Brus, and K. Raghavachari, Chem. Phys. Lett. {\bf 125}, 459 (1986).

\bibitem{Nik}
A.V. Nikolaev and K.H. Michel, Solid State Commun. {\bf 117}, 739 (2001).

\bibitem{Mic2}
K.H. Michel, J.R.D. Copley, and D.A. Neumann, Phys. Rev. Lett. {\bf 68}, 2929 (1992).

\bibitem{Dav}
W.I.F. David, R.M. Ibberson, T.J.S. Dennis, J.P. Hare, and K. Prassides, Europhys. Lett. {\bf 18}, 219 (1992);
{\bf 18}, 735 (addendum) (1992).

\bibitem{Nik2}
A.V. Nikolaev, K. Prassides, and K.H. Michel, J. Chem. Phys. {\bf 108}, 4912 (1998).

\bibitem{Bra}
C.J. Bradley and A.P. Cracknell {\it The Mathematical Theory of Symmetry in Solids}, (Clarendon, Oxford, 1972).

\bibitem{Swi}
T.M. de Swiet, J.L. Yarger, T. Wagberg, J. Hone, B.J. Gross, M. Tomaselli, J.J. Titman, A. Zettl, and M. Mehring, Phys. Rev. Lett. {\bf 84},
717 (2000).

\bibitem{Dav2}
W.I.F. David, R.M. Ibberson, J.C. Matthewman, K. Prassides, T.J.S. Dennis, J.P. Hare, H.W. Kroto, R. Taylor, and
D.R.M. Walton, Nature (London) {\bf 353}, 147 (1991).

\bibitem{Mic3}
K.H. Michel and J. Naudts, Phys. Rev. Lett. {\bf 39}, 212 (1977); J. Chem. Phys. {\bf 67}, 547
(1977).

\bibitem{Lyn}
R.M. Lynden-Bell and K.H. Michel, Rev. Mod. Phys. {\bf 66}, 721 (1994).

\bibitem{Pei}
R.E. Peierls, {\it Quantum Theory of Solids} (Clarendon, Oxford, 1974).

\bibitem{Mic4}
K.H. Michel and J.M. Rowe, Phys. Rev. B {\bf 32}, 5818 (1985).

\end{thebibliography}
\end{document}